\documentclass[journal]{IEEEtran}

\usepackage{xcolor}
\usepackage{grfext}
\usepackage{kvoptions}
\usepackage{kvsetkeys}
\usepackage{adjustbox}
\usepackage{collectbox}
\usepackage{ltxcmds}
\usepackage{cite}
\usepackage{amsmath,amssymb}
\usepackage{textcomp}
\usepackage[normalem]{ulem}
\usepackage{epstopdf}
\usepackage{subcaption}
\definecolor{mycolor}{RGB}{0, 0, 0}

\begin{document}

		\title{Motion Classification Based on Harmonic Micro-Doppler Signatures Using a Convolutional Neural Network}

	\author{Cory Hilton,~\IEEEmembership{Graduate Student Member,~IEEE}, Steve Bush, Faiz Sherman, Matt Barker,\\Aditya Deshpande, Steve Willeke, and Jeffrey A. Nanzer,~\IEEEmembership{Senior Member,~IEEE}
		\thanks{Manuscript received 2022.}
		\thanks{This work was supported by the Procter \& Gamble Company. \textit{(Corresponding author: Jeffrey A. Nanzer)}}
		\thanks{S. Bush, F. Sherman, M. Barker, A. Deshpande, and S. Willeke are with the Procter \& Gamble Company, Cincinnati, OH 45202 USA.}
		\thanks{C. Hilton amd J. Nanzer are with the Department of Electrical and Computer Engineering, Michigan State University, East Lansing, MI 48824 USA (email: hiltonc2@msu.edu, nanzer@msu.edu).}
	}

	\maketitle
	
	\begin{abstract} 
		
	We demonstrate the classification of common motions of held objects using the harmonic micro-Doppler signatures scattered from harmonic radio-frequency tags. Harmonic tags capture incident signals and retransmit at harmonic frequencies, making them easier to distinguish from clutter. We characterize the motion of tagged handheld objects via the time-varying frequency shift of the harmonic signals (harmonic Doppler). With complex micromotions of held objects, the time-frequency response manifests complex micro-Doppler signatures that can be used to classify the motions. We developed narrow-band harmonic tags at 2.4/4.8 GHz that support frequency scalability for multi-tag operation, and a harmonic radar system to transmit a 2.4 GHz continuous-wave signal and receive the scattered 4.8 GHz harmonic signal. Experiments were conducted to mimic four common motions of held objects from 35 subjects in a cluttered indoor environment. A 7-layer convolutional neural network (CNN) multi-label classifier was developed and obtained a real time classification accuracy of 94.24$\%$, with a response time of 2 seconds per sample with a data processing latency of less than 0.5 seconds.
	\end{abstract}

	\IEEEpeerreviewmaketitle


	\section{Introduction}
	
	Increasingly rapid developments in human-computer interaction, the internoet of things (IoT), home health, and other areas is leading to more demand for accurate detection and estimation of the motions of people in living spaces.
	Accurate classification of human motions enables wireless control of devices for interactive entertainment technologies such as augmented reality and virtual reality (AR/VR), determination of human activities to improve home health monitoring, and also differentiation between people and other moving objects. 
	Classifying the motions of people in living spaces is challenging, however, due to the complexity of the environment. Optical systems may be used, however they often entail computationally expensive image processing and have privacy concerns since images of people are formed and processed. Infrared systems generally operate under similar principals and have similar drawbacks. 
	Microwave radar has been increasingly used in short-range motion detection applications due to a number of beneficial aspects that radar holds over optical systems~\cite{6697945,8755821}. Radar systems process complex signal returns, so that motion can be directly measured via Doppler frequency shifts, unlike optical systems which are generally intensity-based, requiring techniques like change detection to infer motion. Microwave radar also provides a reliable method of detecting moving objects without forming images of people, alleviating such privacy concerns. Furthermore, when monitoring the motions of objects with rapid variations, such as human movements, the micromotions of the body generate distinguishable signatures in the time-frequency response of the radar return, called micro-Doppler signatures, which can be used for classification of the micromotions~\cite{4801689,7444065,8944152,8970277,9130759}. 
	
	One of the principal challenges in detecting and classifying micro-Doppler signatures in living spaces is the presence of significant signal reflections from the environment. These clutter returns can be significantly higher in power than that returns from a moving person. While the clutter returns are centered at zero Doppler, the phase noise of the radar system contributes to a broader frequency response that overlaps with slow moving targets making them indistinguishable. Since human motions tend to be relatively slow, these micro-Doppler responses can thus be challenging to detect in highly cluttered indoor environments, and even if detected, the presence of system noise may reduce the classification accuracy.
	Harmonic radars and harmonic tags have been developed to overcome these challenges by moving the signal response to a frequency band without clutter\cite{o2004tracking,colpitts2004harmonic,bottigliero2019innovative,hayvaci2021linear}. A harmonic tag accepts the incident radar transmit signal at the fundamental frequency $f_c$ and using a nonlinear component such as a diode generates a higher harmonic signal. Typically the second harmonic $2f_c$ which is then retransmitted and captured by the radar receiver. Since all the clutter returns are present only near $f_c$, the harmonic tag signal can theoretically be detected more easily, however the conversion efficiency of the tag dictates how much return power the radar captures.

		\begin{figure*}[t!]
		\begin{center}
			\noindent
			\includegraphics[width=0.95\textwidth]{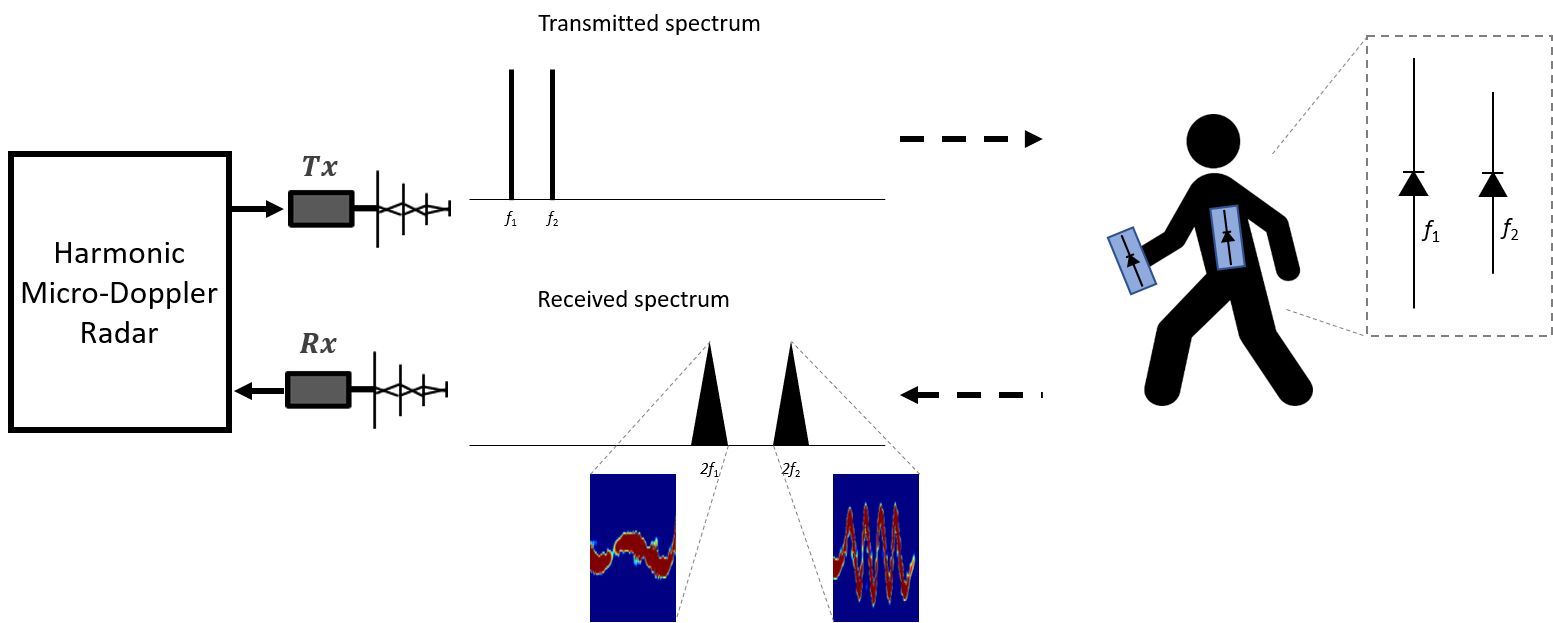}
			\caption{Simultaneous measurement of the motion of multiple tags is enabled by narrow-band tag responses. Tags are tuned to specific resonance frequencies matching a set of narrowly spaced tones transmitted by the radar. The tags generate signals at the second harmonic of the incident signals and retransmit them back to the radar, which can detect the harmonic responses without interference from environmental clutter. The harmonic frequency responses are modulated by the micromotions of the tags on the body, producing harmonic micro-Doppler signatures that can be processed in the time-frequency domain for activity classification.}
			\label{OV-1}
		\end{center}
	\end{figure*}

	While previous works on harmonic tags have principally focused on detection, we recently presented a method of detecting the motion of harmonic tags via the harmonic Doppler return and demonstrated the presence of micro-motion information in harmonic micro-Doppler signatures~\cite{9049125,9366931}. When detecting the motion of multiple objects, differentiability of the separate radar returns is necessary. To address this with a low-cost tag, we recently designed a narrow-band harmonic tag that can be used for frequency-selective harmonic Doppler measurements such that multiple tags can be present in the operational bandwidth of the radar while maintaining sufficient frequency separability to distinguish the motions of multiple tags independently~\cite{9947030}. 
	
	In this paper we demonstrate for the first time the ability to classify multiple human motions using only their harmonic micro-Doppler radar signature. We use narrow-band tags to support scalability and a harmonic radar system operating at 2.4/4.8 GHz. The tag is based on a narrow-band 2.4 GHz split ring antenna design combined with a diode and a simple 4.8 GHz dipole for retransmission at the second harmonic. We collected measurements of four different common hand motions from a set of 35 different participants in a cluttered indoor environment. A 7-layer convolutional neural network (CNN) was designed to classify the four gestures in real time. An average classification accuracy of 94.24\% was obtained, demonstrating the ability to accurately classify the motions of held objects in cluttered indoor environments. The rest of this paper is outlined as follows: In Sections \ref{S-tag} and \ref{S-radar} we describe the design of the narrow-band harmonic tag and the harmonic radar system. The CNN classifier is detailed in Section \ref{S-classifier}. Experimental results demonstrating the classification of tag motions are presented in Section \ref{S-results}.

		\section{Harmonic Tag Design}\label{S-tag}

	
	The objective of the system is to enable detection and tracking of multiple tagged objects simultaneously. Whereas harmonic tags can be embedded with active circuitry to modulate the response of the tag, thereby imparting a unique signature on the response from individual tags. This approach to differentiability requires more complex tags involving additional circuitry and potentially energy harvesting. Our approach is based on differentiability in the frequency domain through narrow-band harmonic tag design (see Fig. \ref{OV-1}). In this approach, the tags are simple and low-cost, consisting of only two antennas and a diode to generate the return signal at a harmonic of the incident signal. The narrow bandwidth is imparted by one of the antennas to minimize the need for filtering circuitry on the tag. We designed the tags with a narrow-band response at the fundamental frequency using a split ring antenna that can easily be tuned for different frequencies. The radar system emits multiple tones that are matched to the resonance frequencies of the ring antennas on the tags, thereby producing responses at the second harmonic of the transmitted signals if the tags are present, facilitating detection. If the tags are present and moving, the second harmonic signal response will also be modulated by the motion of the tag, imparting a micro-Doppler signature on the harmonic signal. The tags are designed for minimal overlap of the resonance bandwidth, which is wider than the expected micro-Doppler frequency modulation, thus the responses can be individually detected at the harmonic radar receiver.
	

	
	Harmonic radar is based on the transmission of a signal at a fundamental frequency and reception of the signal retransmitted by the harmonic tag at a harmonic of the fundamental frequency. The transmitted signal may be represented by 
	\begin{equation}
	S_{TX}(t) = Ae^{-j2\pi f_1 t}
	\end{equation}
	where $A$ is the signal amplitude and $f_1$ is the fundamental frequency. The signal retransmitted by the harmonic tag and received at the radar can be represented by
	\begin{equation}
	S_{RX}(t) = Be^{-j2n\pi f_1 \left(t-\tau(t)\right)}
	\end{equation}
	where $B$ is the amplitude, $n$ is the harmonic number of the retransmitted signal, and $\tau(t) = 2r(t)/c$ is the propagation delay, which varies as a function of time due to the changing distance to the radar $r(t) = r_0 - v_r(t)t$ from the motion of the tag. The signal is thus centered on the harmonic frequency $f_n = nf_1$ and if the tag is moving it will be imparted with a harmonic Doppler frequency shift given by
	\begin{equation}
	f_{D,n}(t) = 2n f_1 \frac{v_r(t)}{c} = \frac{2nv_r(t)}{\lambda_1}
	\end{equation}
	where $\lambda_1$ is the wavelength of the fundamental frequency. The time-varying motion of the tag thus manifests as the time-varying harmonic micro-Doppler frequency, which can be processed to characterize the motion of the tag.
	
	An important metric when considering tag design is its conversion efficiency, i.e., how much power of the signal at the fundamental frequency is converted into the retransmitted power at the harmonic frequency. Other important design metrics of the tag are the gain of the antennas at the fundamental and harmonic frequencies. These are captured in the harmonic scattering cross section of the tag, given by~\cite{9049125}
	\begin{equation}
		\sigma^{h} = \epsilon_{n} \frac{\lambda^{2}_{1}}{4\pi} G_{tag,1}G_{tag,n}
	\end{equation} 
where $\epsilon_{n}$ is the conversion efficiency at the $n$th harmonic, and $G_{tag,n}$ is the gain of the tag at the $n$th harmonic. The harmonic scattering cross section is used in the harmonic radar range equation which characterizes the received power at the $n$th harmonic by~\cite{9049125}
\begin{equation}
p_{r,n} = p_{t,1} \frac{G_{t,1} G_{r,n}}{(4 \pi)^{3} r^{4}} \lambda _{n}^{2} \sigma ^{h} \tag{8} \end{equation}
where $p_{t,1}$ is the transmitted power at the fundamental frequency, $G_{t,1}$ is the gain of the radar transmit antenna at the fundamental frequency, $G_{r,n}$ is the gain of the radar receive antenna at the $n$th harmonic frequency, and $\lambda _{n}$ is the wavelength of the signal at the $n$th harmonic frequency. While the radar in this paper uses a continuous-wave signal, the received power may be increased by using pulsed radar to both increase the peak power and to support coherent pulse integration~\cite{9366931}.

\begin{figure}[t!]
		\centering
		\includegraphics[width=.5\textwidth]{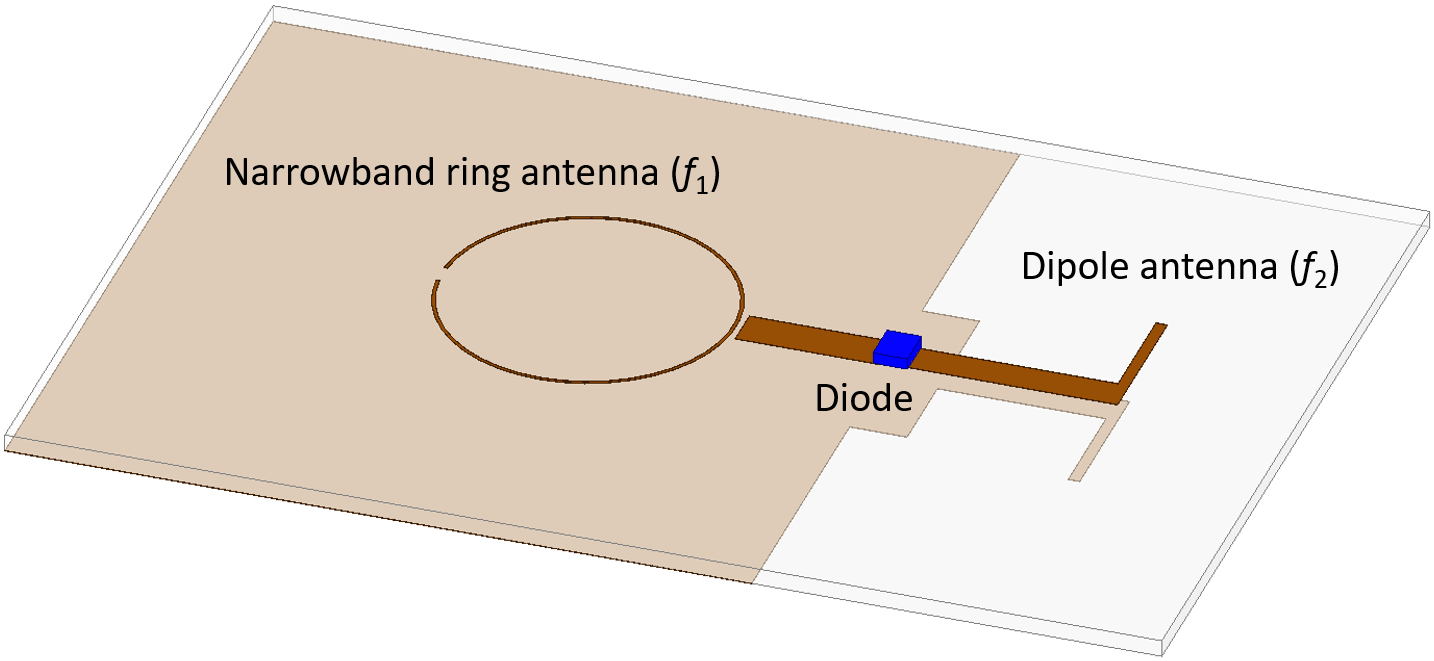}
		\caption{Diagram of the harmonic tag consisting of a narrow-band split ring antenna resonant at the fundamental frequency, a diode to generate harmonics, and a dipole resonant at the second harmonic~\cite{9947030}.}
		\label{antenna}
	\end{figure}

The harmonic tag is based on a narrow-band split ring antenna designed to resonate at the fundamental frequency (see Fig. \ref{antenna})~\cite{9947030}. The output of the ring antenna is passed through a diode (BAT15-03W) which generates harmonics of the incident signal. The output of the diode is connected to a simple dipole antenna which resonates at the second harmonic frequency. The ring antenna provides a narrow-band response, the center frequency of which is designed by changing either the radius of the ring or the gap width. The dipole is sufficiently wideband that the same design can be used for multiple tags within the same operating bandwidth, thus multiple tags in the scene can be designed by changing only the split ring antenna.

The tag was designed using ANSYS HFSS and manufactured on 1.524 mm Rogers 4350B with a 39 $\mu$m copper cladding. The ring antenna had a gain of $G_{tag,1} = 2.1$ dBi and a bandwidth (measured at the -10 dB S11 points) of 11 MHz at a resonance frequency of 2.39 GHz, for a fractional operational bandwidth of 0.46\%. The measured beamwidth of the antennas covered the front-facing hemisphere with negligible degradation except at grazing angles. Measurements using the harmonic radar described below yielded a conversion efficiency of $\epsilon_2 = 1.6\%$ and a harmonic scattering cross section of 0.61 cm$^2$ for the tags, comparable with that of other works~\cite{colpitts2004harmonic}.

		\begin{figure*}[t!]
		\begin{center}
			\noindent
			\includegraphics[width=0.8\textwidth]{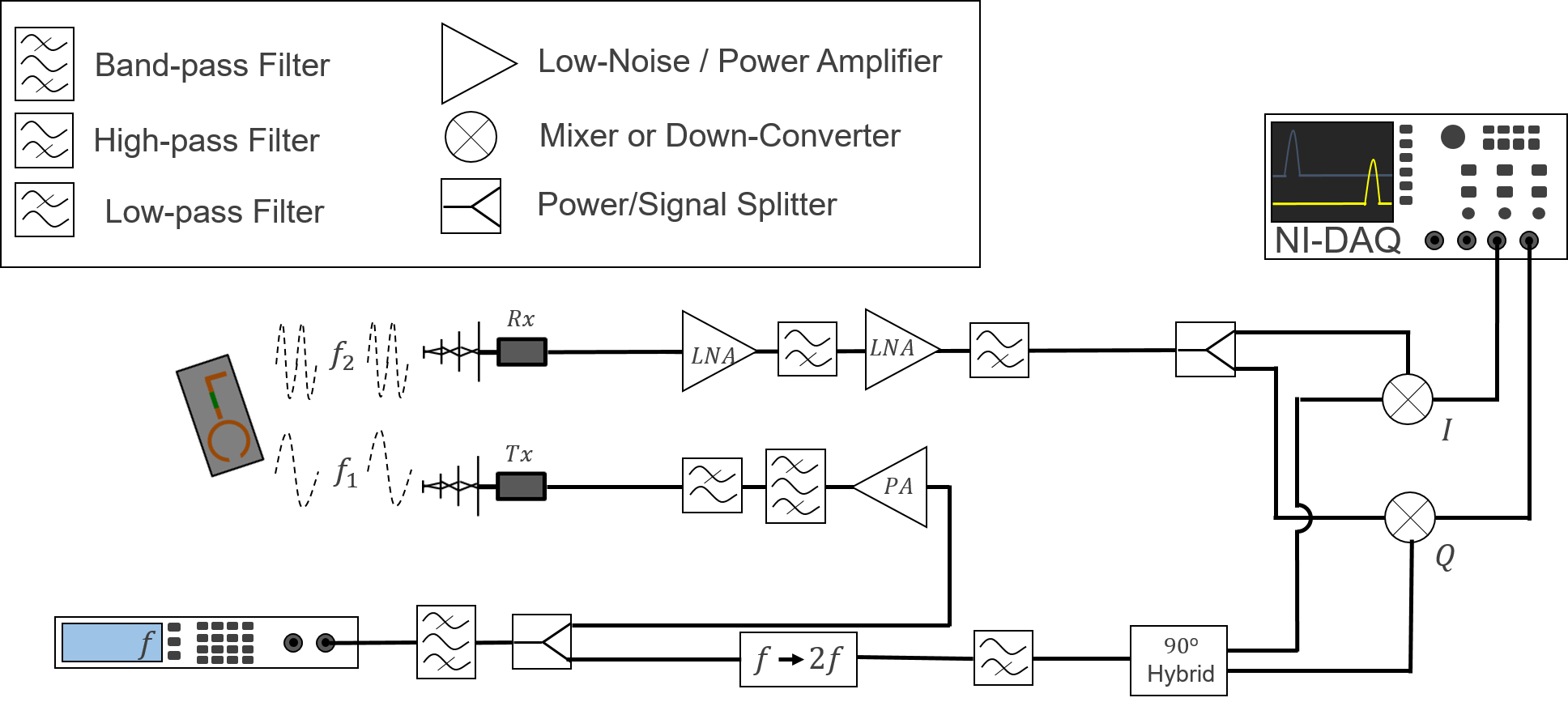}
			\caption{Block diagram of the harmonic radar system. Filtering is used heavily throughout the radar system to ensure that signals existing at harmonics of the transmit frequency are mitigated in the transmitter and that any signals remaining at the fundamental frequency are mitigated in the receiver. This is to ensure that the received signals are due only to the retransmissions from the tags.}
			\label{radar}
		\end{center}
	\end{figure*}

		\section{Harmonic Radar System}\label{S-radar}
		
\subsection{Radar Design}

The block diagram of the radar is shown in Fig. \ref{radar}. The radar was designed to transmit in the 2.4 GHz fundamental frequency band and to receive retransmitted signals at the second harmonic frequency at 4.8 GHz. The signal may be modulated to transmit a set of tones as shown in Fig. \ref{OV-1}. However, for this work we focus on the the classification of single tags, thus only a single frequency is used. The radar consists of a local oscillator (LO) signal generated at 2.4 GHz with a power of 6 dBm by a Keysight N5183A signal generator that is split two ways, with one signal fed to the transmitter and the second to the harmonic receiver. The transmit signal is first amplified to boost the transmit power using a 29-dB gain amplifier (Mini-Circuits ZX2700MLNW+), after which a series of filters are used to mitigate the transmission of any residual second harmonic signals. It is important that any signals at the second harmonic frequency generated by the LO or the amplifier are highly suppressed to ensure that any signal received by the radar at the second harmonic are generated only by the tag, and are not transmitted and scattered off the scene. 
The transmit and receive side of the system used identical log-periodic antennas with bandwidths of 2.3--6.5 GHz and a typical gain of $8$ dBi.

The signal received by the radar was amplified using two 22 dB gain low-noise amplifiers (Mini-Circuits ZX60-542LN-S+), each followed by high-pass filters (Mini-Circuits ZFHP-3800FF-S+) to mitigate any residual fundamental frequency signals either scattered off the scene or coupled between the antennas. The received signal was then split for quadrature downconversion.
The second output of the LO power splitter was input to a frequency doubler (Mini-Circuits ZX90-2-36-S+) which generated the second harmonic of the incident signal at 4.8 GHz. The output of the doubler was filtered to remove and residual signal at the fundamental frequency. This signal was input to a hybrid coupler (RF-Lambda RFHB02G08GVT) to generate a quadrature LO signal for the downconverter. The outputs of the coupler were input to the LO ports on two frequency mixers (Mini-Circuits ZX05-14-S+).  
The in-phase (I) and quadrature (Q) mixer outputs were then collected using a NI-9324 USB DAQ in conjunction with the Python DAQmx API.
Performing a cascade analysis of the receiver chain and using the harmonic radar range equation and the harmonic scattering cross section of the tags, the SNR of the received signal was approximately 16 dB for typical measurements.

			\begin{figure}[t!]
		\centering
		\includegraphics[width=.5\textwidth]{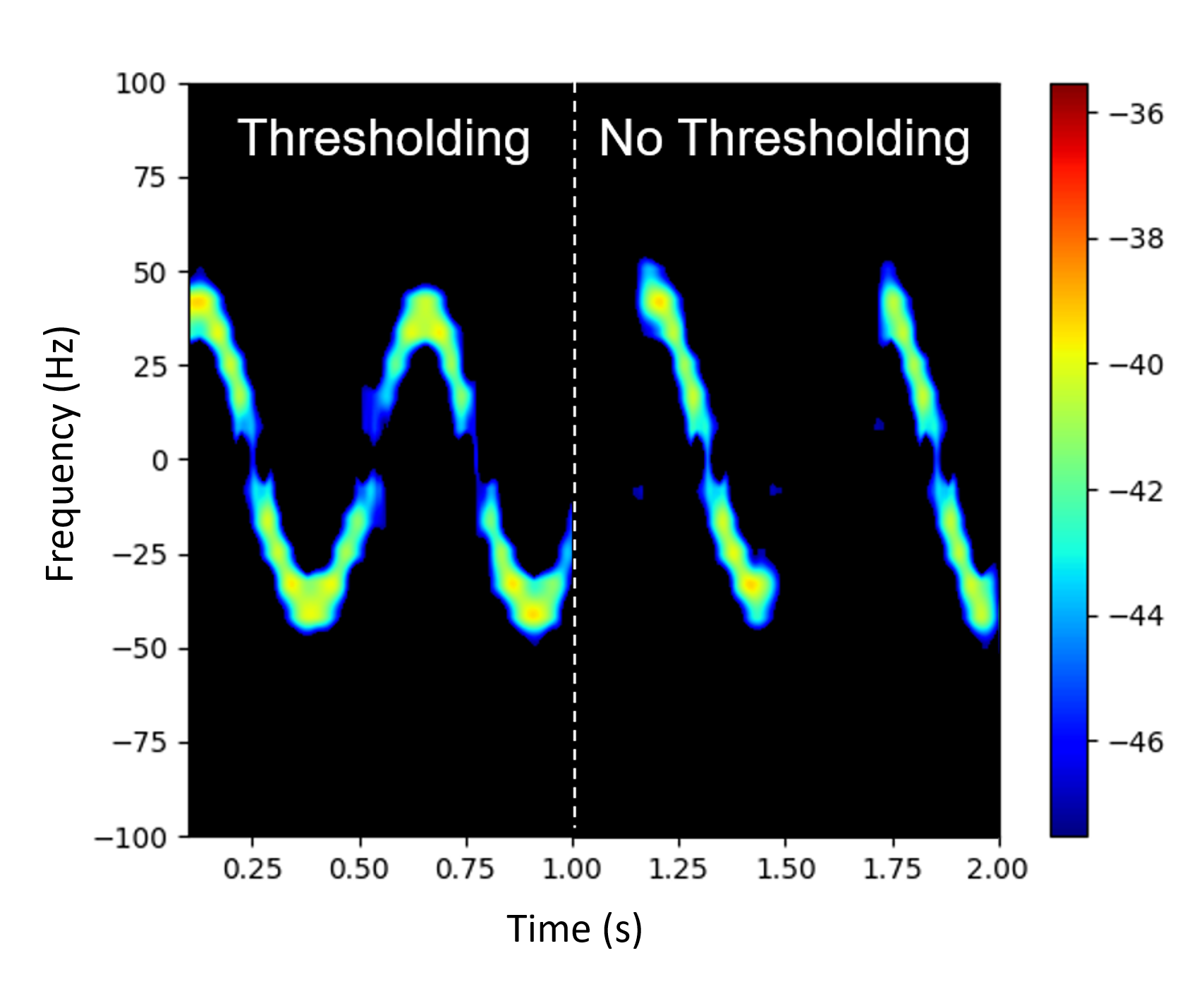}
		\caption{Example spectrogram showing the change in dynamic range with and without thresholding.}
		\label{detrend}
	\end{figure}
	
	\begin{figure}[t!]
		\begin{tabular}{cc}
			\includegraphics[width=0.45\linewidth]{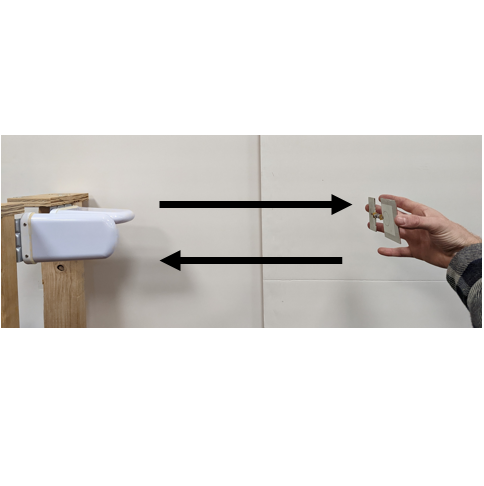} &   \includegraphics[width=0.45\linewidth]{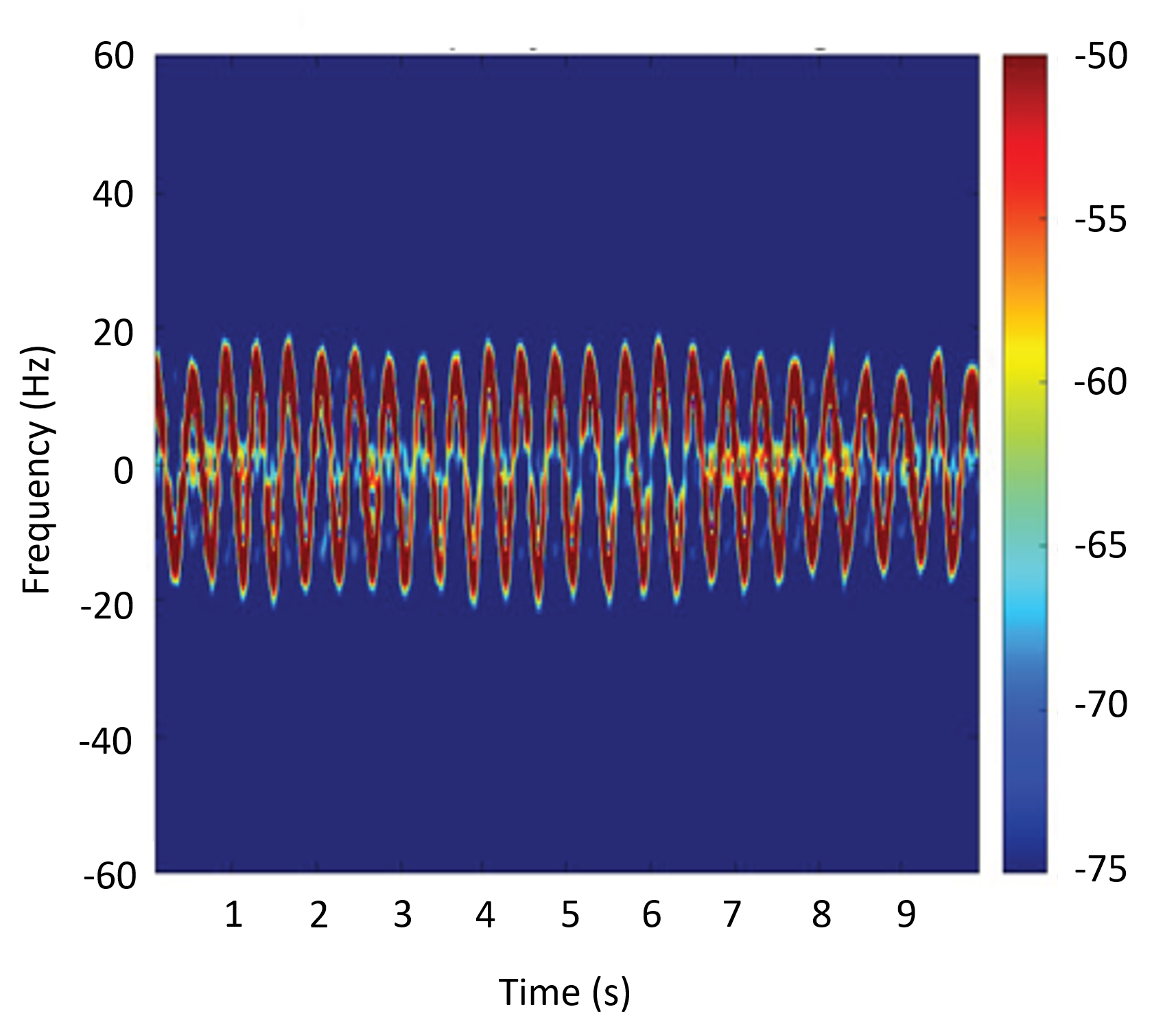} \\
			(a)  & (b)  \\[6pt]
			\includegraphics[width=0.45\linewidth]{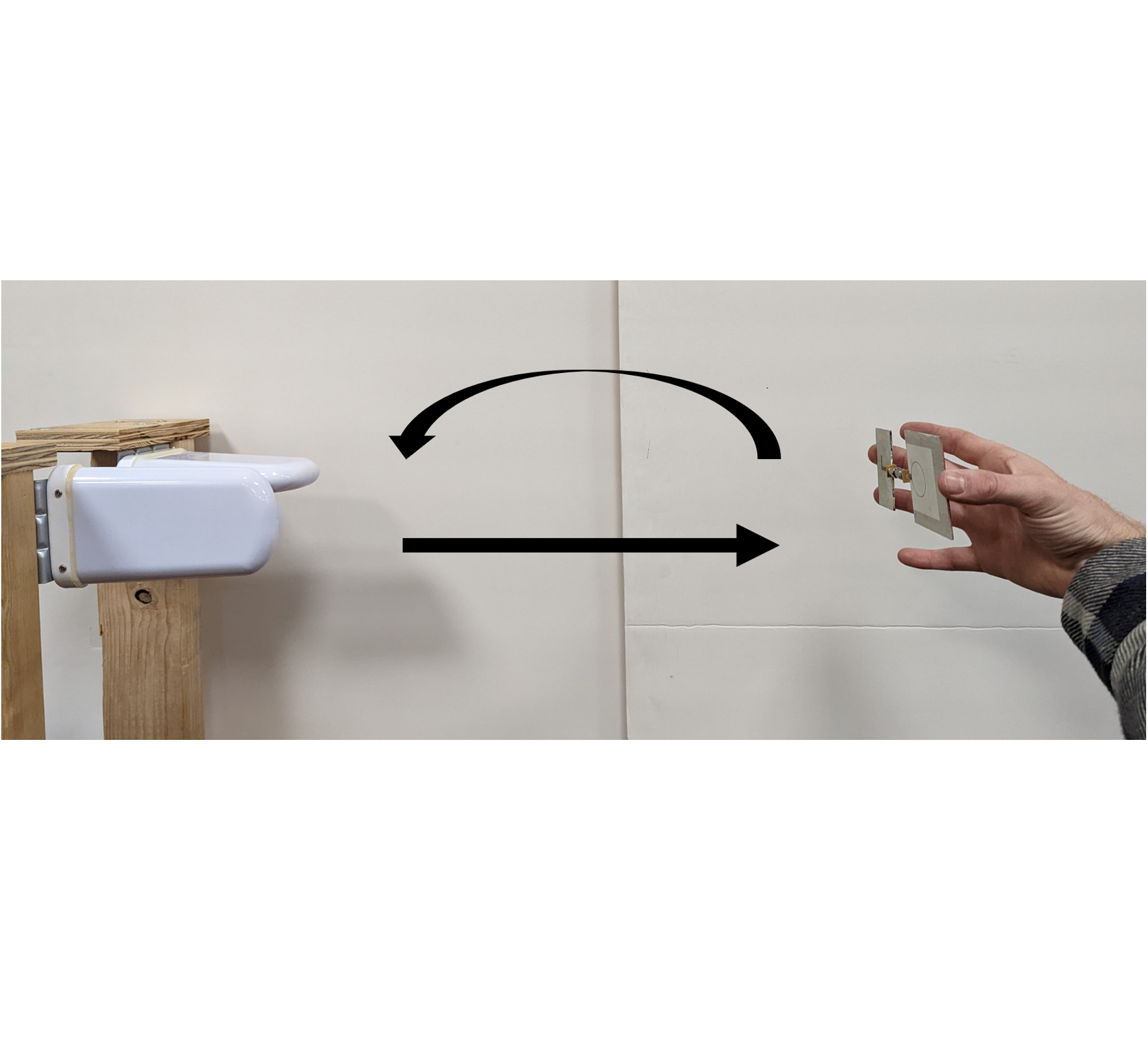} &   \includegraphics[width=0.45\linewidth]{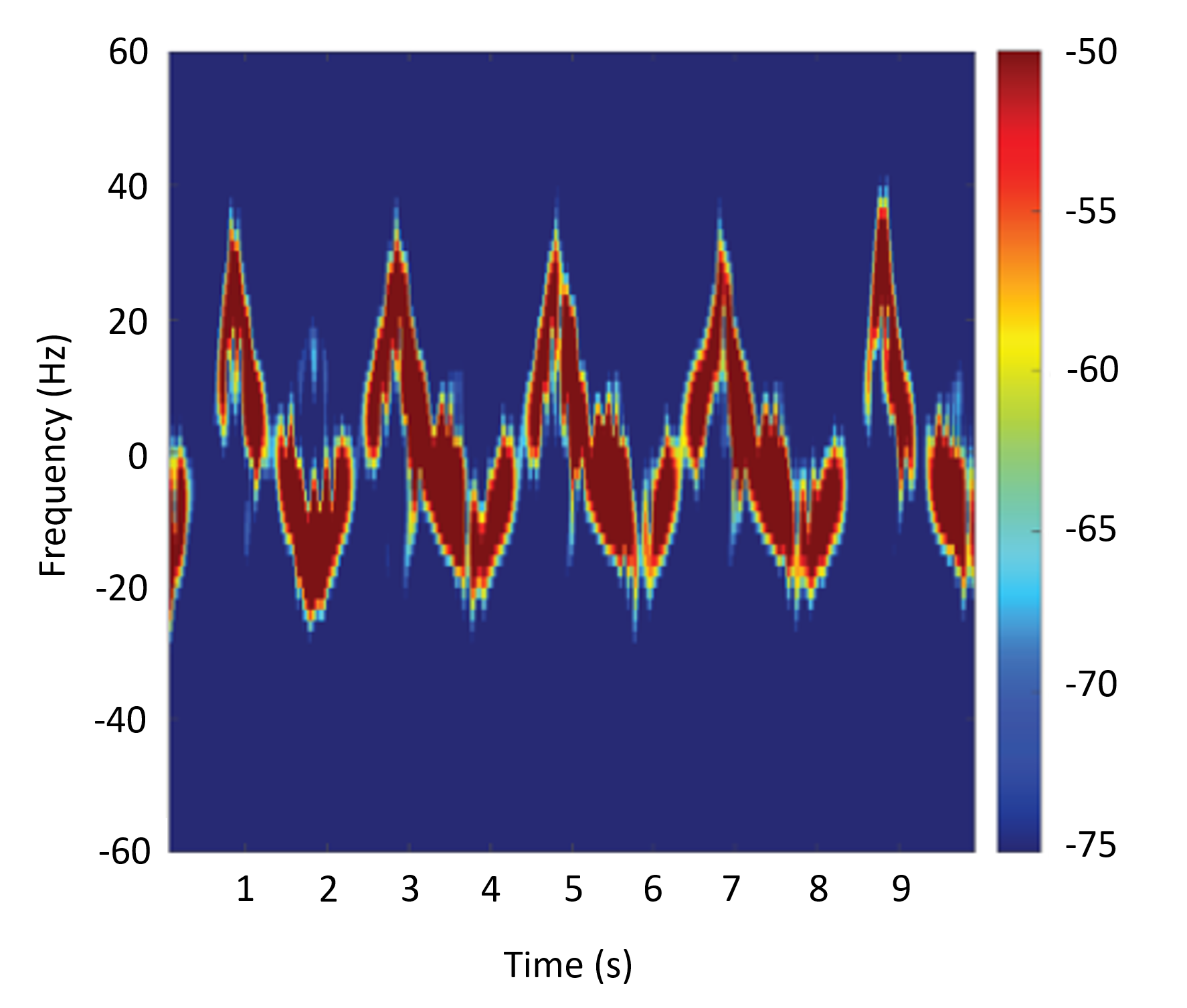} \\
			(c)  & (d)  \\[6pt]
			\includegraphics[width=0.45\linewidth]{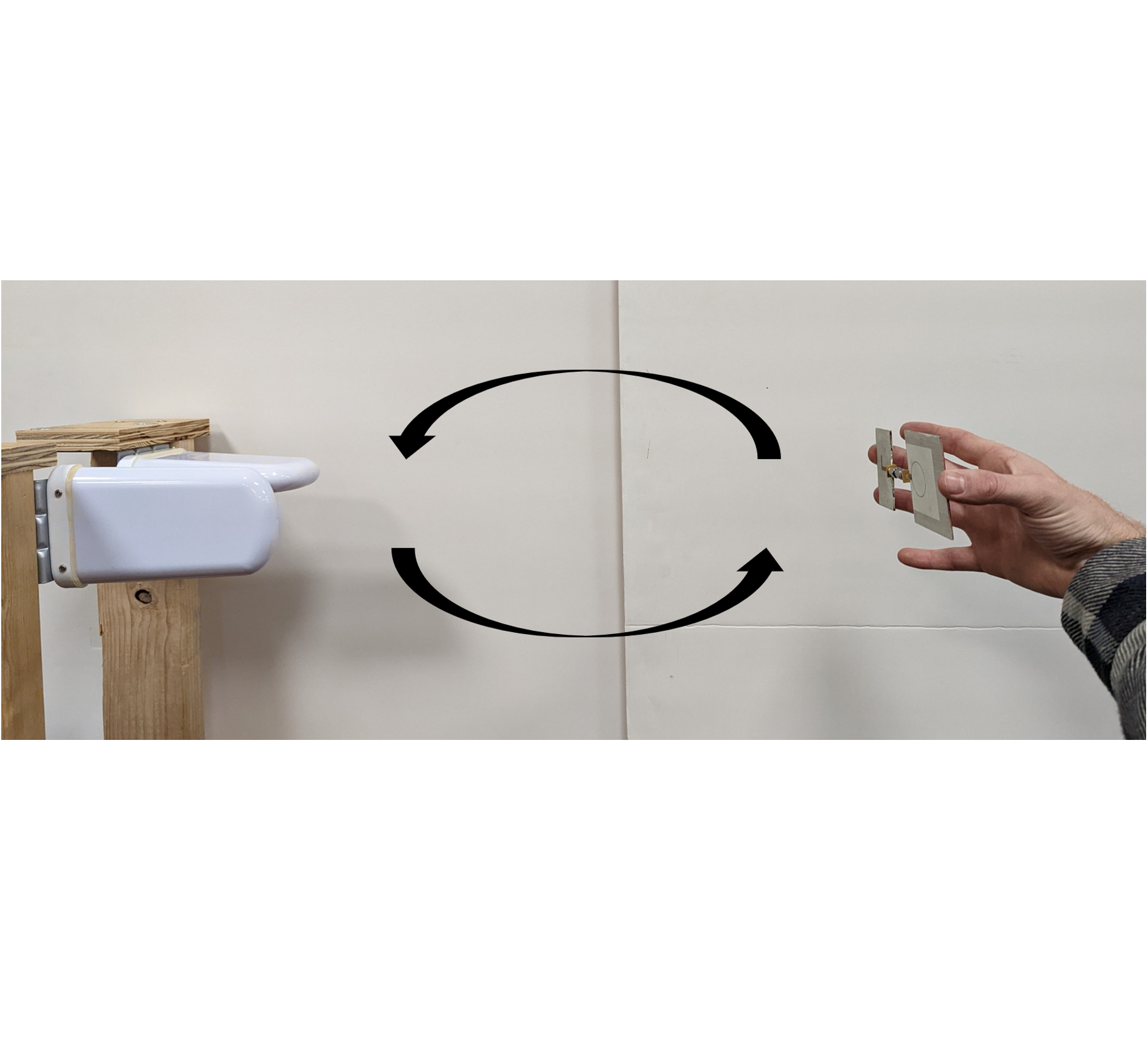} &   \includegraphics[width=0.45\linewidth]{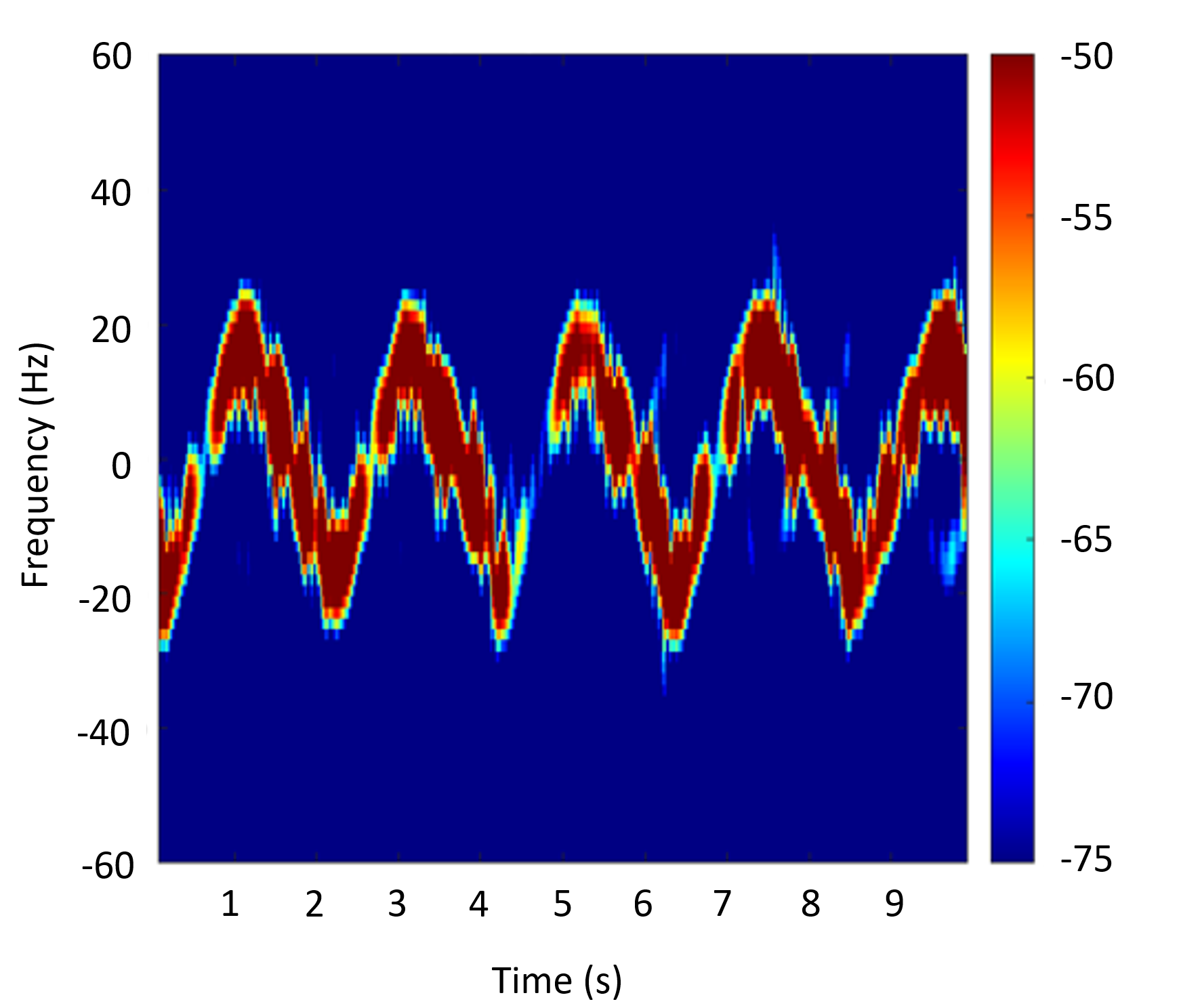} \\
			(e)  & (f)  \\[6pt]
		\end{tabular}
	\caption{Activity motions used for classification of the harmonic Doppler radar responses and the corresponding motions associated with each classification. Activity motion 1 (a,b) shows a fast ($>2$ m/s) periodic short distance motion performed within the beamwidth of the log-periodic antennas. Activity motion 2 (c,d) shows a fast motion followed by a slow motion over a 0.5-1m distance performed within the beamwidth of the log-periodic antennas. Activity motion 3 (e,f) shows a 0.5-1 m/s periodic motion over a 1m distance performed within the beamwidth of the log-periodic antennas.}
	\label{motions}
	\end{figure}

	\begin{figure*}[t!]
		\begin{center}
			\noindent
			\includegraphics[width=0.75\textwidth]{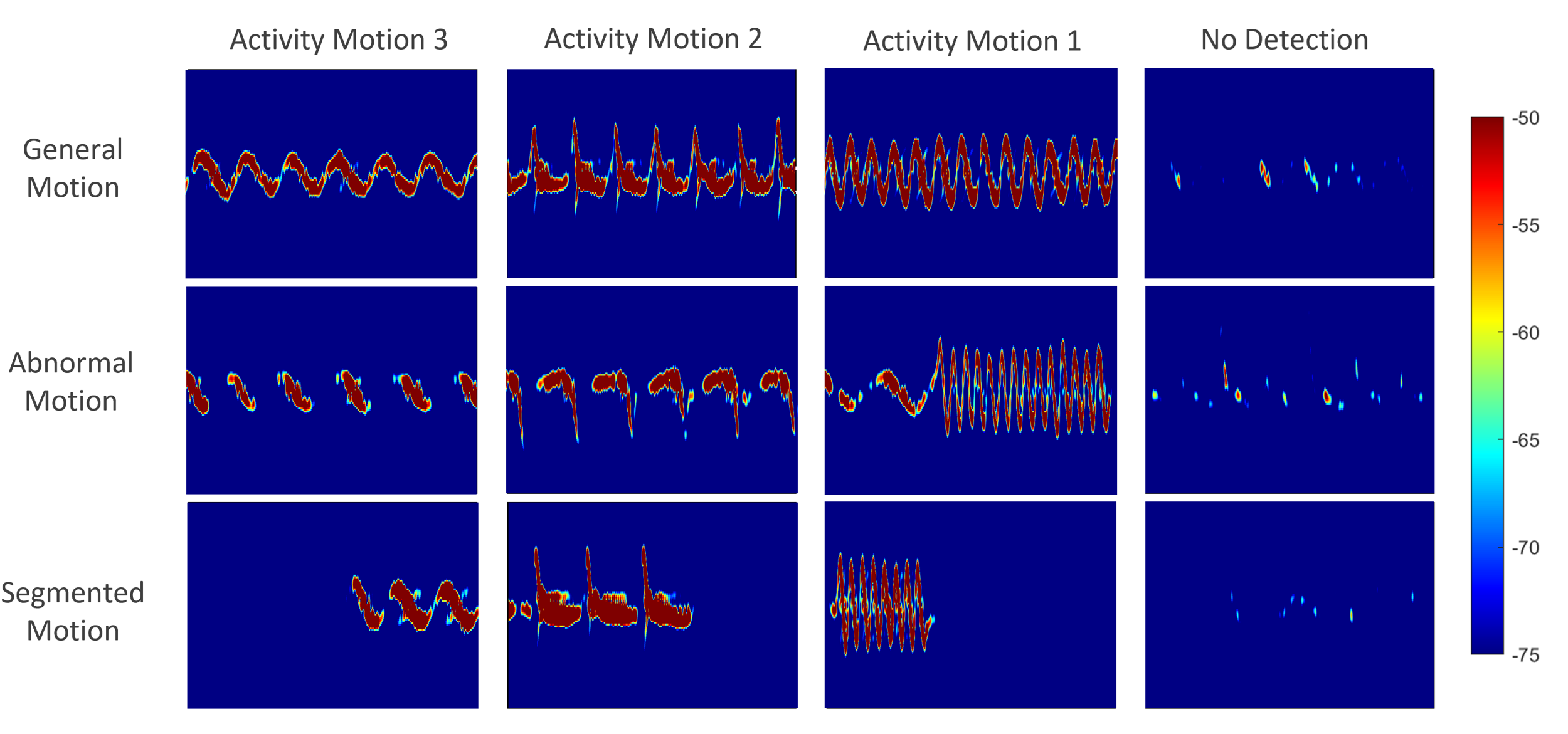}
			\caption{Short time spectrogram captures of varying methods of performing designated activity motions collected from the harmonic Doppler radar and manually classified.}
			\label{ov}
		\end{center}
	\end{figure*}
	
\begin{figure*}[]
		\centering
		\includegraphics[width=0.75\linewidth]{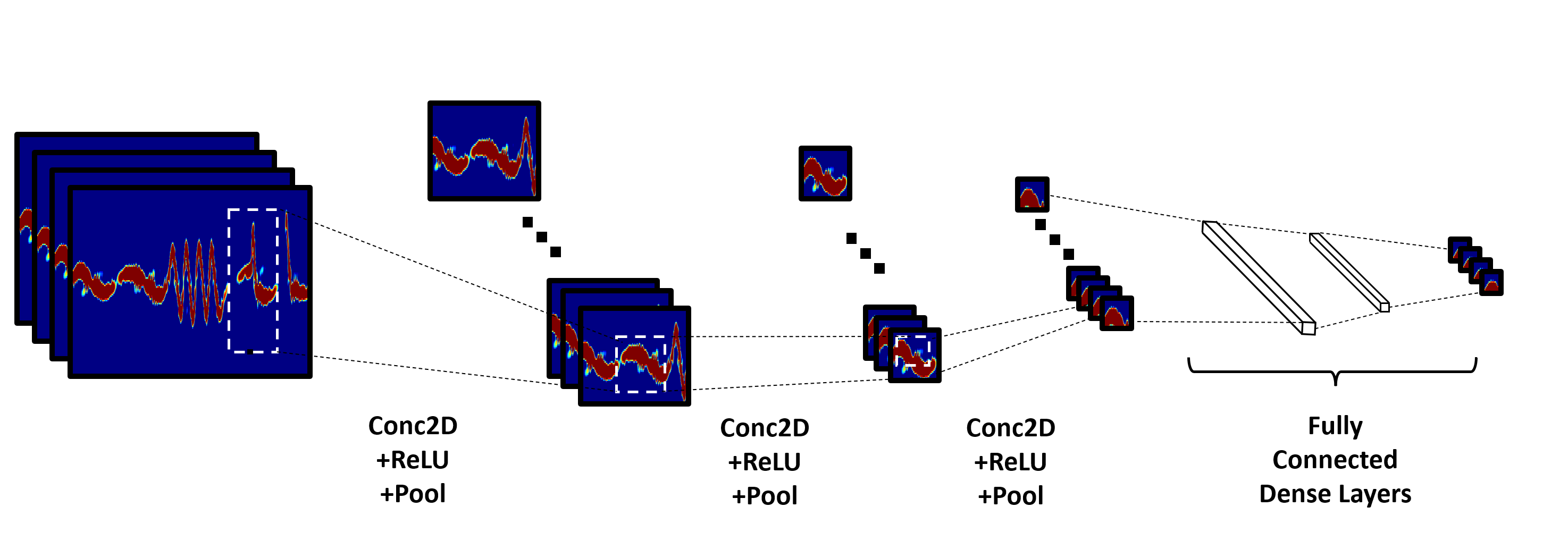}
		\caption{Model of the multi-label CNN  classifier wherein the training model is comprised of six 2D convolution layers with each followed by a max pooling and dropout layer respectively to combat the over-training common to this project. These were then followed with four fully connected dense layers allowing the network to estimate any given image to one of the four provided classes. This network was designed using the Tensorflow package on Python 3.10 with the Adam optimizer providing a learning update of 0.001 and a betas default value of 0.99. }
		\label{cnn}
\end{figure*}
	
\subsection{Signal Processing}

	The I/Q voltage signals collected from the NI-9234 DAQ were processed using Python with a sample rate of 2.5 kSa/s. The expected harmonic Doppler shifts are on the order of 60 Hz for typical held object motion, thus the sample rate was more than sufficient to capture the full harmonic micro-Doppler spectrum without aliasing.
	The signals were processed initially to correct for residual I/Q phase mismatch, and then processed using a detrending algorithm and a $5^{th}$ order Butterworth low pass filter to remove any residual dc bias or low-frequency noise on the received signals. 
	The signals were analyzed using the short-time Fourier transform (STFT) which plots the spectrum of the signal over time using a sliding window. The spectrogram of the signal (the magnitude of the STFT) was then generated and used in the classifier. 
	The spectrograms were also processed using a threshold to constrain the dynamic range and generate spectrogram images with more information. An example response of a sinusoidal radial motion with and without thresholding is shown in Fig. \ref{detrend}.
	
	In Fig. \ref{motions} three different motions of the tag in front of the radar, along with the their corresponding harmonic spectrograms. The micromotion of the tags clearly manifests in the harmonic micro-Doppler signatures. Differences in the time-frequency responses of different motions support classification. For example, periodic motion of the tags generates periodic responses in the harmonic spectrogram; differences in the periodicity of the motions clearly manifests in the spectrogram. Furthermore, motions with similar periodicities (see Figs. \ref{motions}(c) and (e)) but different trajectories yield differences in the response: The motion in Figs. \ref{motions}(c) produces a time-frequency signal that is peakier (higher kurtosis) that the response of Figs. \ref{motions}(e) which is a smoother sinusoid (lower kurtosis). These differences support robust classification of the motions.

	\section{Convolutional Neural Network Classifier}\label{S-classifier}
	
	Convolutional neural networks have seen a large increase in usage for RF and Doppler classification as well as many image processing applications due to a few key features they possess as compared to competing methods of feature extraction \cite{lambooi,DNN,chenradar}. These include invariance to translations, illumination, or other noise factors commonly associated with these responses. In this work we applied the CNN to the time-frequency responses of a single receiver Doppler output wherein the results were then compared to the four activity motion classes. 	
	
	Spectrograms of the four classes are shown in Fig. \ref{ov}, with the three motions as described in Fig. \ref{motions} and a fourth class that is no motion detected. Fig. \ref{ov} shows examples of various cases for each class: a general case showing a controlled motion, an abnormal motion case where the motion was, e.g., changed mid-motion, and segmented motion, where only a partial instance of the motion was captured. 
	
A multi-label convolutional neural network was designed to classify the spectrograms based on the theory of general image classification. The classifier was designed to accept constant updates of the spectrogram images support real time classification. The network topology is shown in Fig. \ref{cnn}. The feature extraction for this classifier was performed by a series of six two-dimensional convolutional layers each followed by a single max pooling layer of a 3x3 pixel matrix as well as a 25\% dropout layer. This was chosen because the total image size for these test images was reduced to 300x300 pixels for each case. The output classification was them performed via four fully connected dense layers yielding and an estimated number of 6,803,000 trainable parameters. In each case the network was trained with a multi-label cross entropy loss function using stochastic gradient descent with a learning rate of adam $=0.001$ \cite{adam} with a default betas value of 0.99. The network was implemented on Python 3.10 with the use of the Tensorflow library. This proved to be a novel method of motion classification as this model was able to perform real time classification on a mobile Intel i5-1145G7 chipset with no dedicated GPU.
	
	Data were collected from two sets of participants: one set of five people who were trained on each motion by being shown the motion beforehand; and one set of 30 participants who received only a brief description of the motion. Three cases were considered: Case 1 consisted of only the five trained participants; Case 2 consisted of only the untrained participants; and Case 3 consisted of all participants. The data were divided into training and evaluation sets as follows: Case 1 used a training data set of 7,002 samples and an evaluation data set of 1,140 samples; Case 2 used a training set of 3,120 samples and an evaluation set of 379 samples; Case 3 combined the two cases for a training set of 10,122 samples and an evaluation training set of 1,519 samples. Each sample consisted of a two-second spectrogram image.
	
	\section{Classification Results}\label{S-results}
	
		The loss function used in this work is defined as
	\begin{equation}
		E = \Sigma^{n}_{0}y^{n}_{true}\log_{10}(y^{n}_{predicted})
		\label{loss}
	\end{equation}
where $y_{true}$ and $y_{predicted}$ refer to the values of the true and predicted classes which have a value between on and four. The summation of these values provides a complete loss function across all images in the training process.
		This allows the implementations of thresholding techniques on all classes to be made wherein the summation of all classes is now a total possible value of four as opposed to the traditional value of one. This enables the implementation of a 5th hidden class in which the summation of any two classes that exceeds a threshold value of 1.5-1.75 will register as a multi-class unrecognized response which can then be added to activity class four for simplicity. 
		For the given data sets, the loss function begins to increase after 5 and 18 epochs for Cases 2 and 3. This could be considered to be the point of over-fitting, however, the chosen loss function contains the multi-label addition not common to the traditional cross-entropy model as well as a nominally small loss value across all epochs in the Case 3 data set. This indicates that increasing the dataset would also yield an improved, lower evaluation loss value as the ambiguity in repetitive human motion provides large shifts from varying subjects performing similar motions. In the following, the best classification accuracy was obtained for Cases 1 and 3 using 60 epochs and for Case 2 using 50 epochs.

			\begin{figure}[t!]
		\centering
		\includegraphics[width=.5\textwidth]{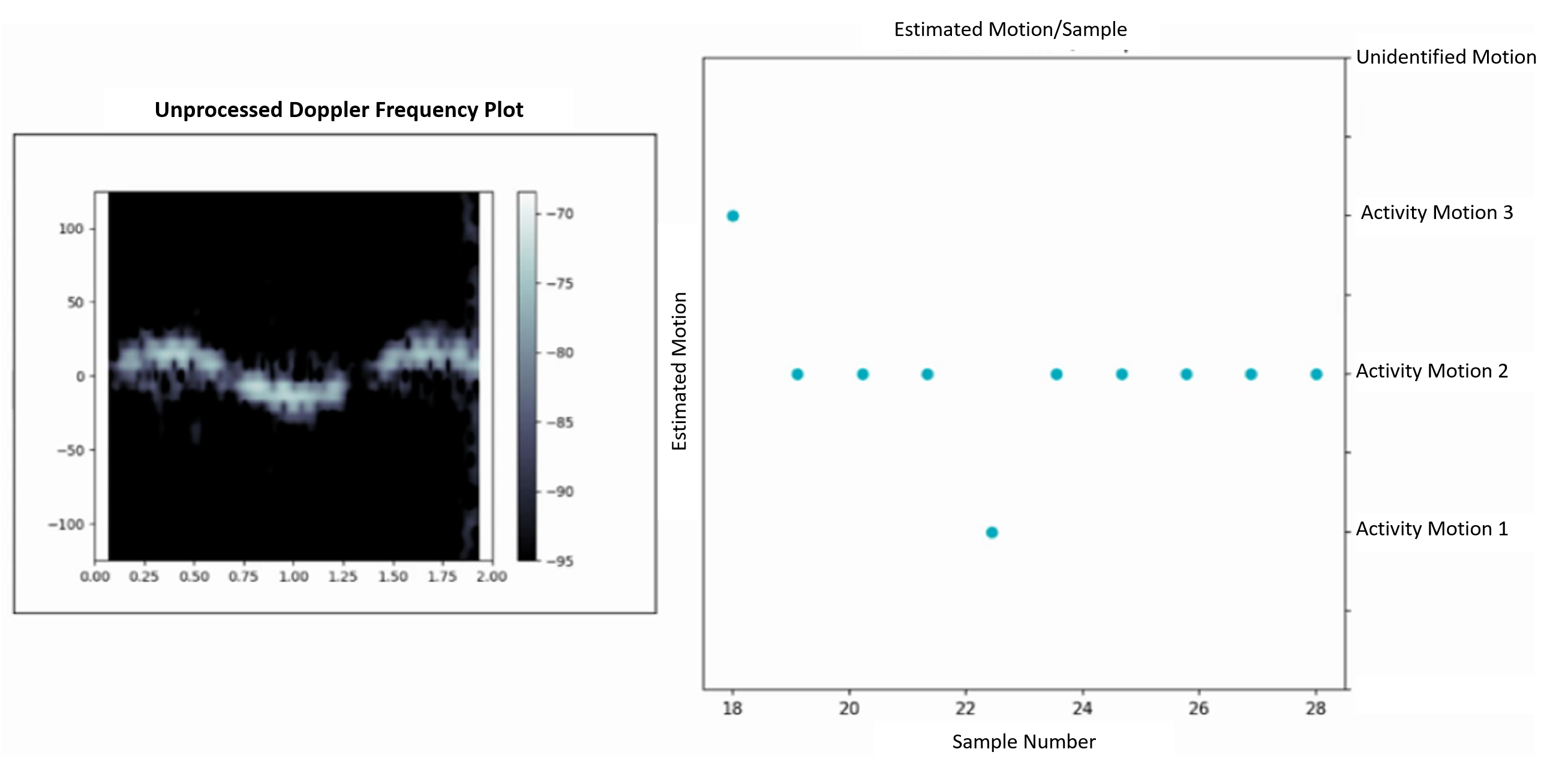}
		\caption{Real-time monitoring GUI allowing for up to 3 fps monitoring of the motion of harmonic tags in cluttered environments. The system plots the spectrogram and outputs the current estimated class of motion.}
		\label{GUI}
	\end{figure}
	
	The classifier was implemented to support real time classification performance. 
	The processing used a thresholding technique for a continuously monitoring the received signal baseline power response, allowing the the radar system to be moved in relation to the tags with little loss of measurement accuracy while within the operational range of about 2 m. The threshold was updated every 10 measurements to estimate the received power of the current activity class, supporting a more consistent dynamic range in the resultant spectrograms. 
	The output spectrogram was represented as a binary image that was thresholded to filter undesired Doppler responses.
		The classifier accepts the binary images of the micro-Doppler responses and plots the results of the estimation. The estimated class was also saved for later re-training. The graphical use interface (Fig. \ref{GUI}) updated at a rage of 3 frames per second.
	

	
	The classification accuracy is shown in Table \ref{res1}, where consistently high accuracy was obtained for all cases.
	The case with trained participants yielded nearly perfect results, however even the more relevant case of untrained participants yielded high accuracy of at least 93\%. 
	The dataset including all participants yielded an average classification accuracy of 94\%, comparable to the untrained dataset. The slight reduction in performance indicates that some over-fitting may be present, or that the differences between the trained and untrained actions are enough to introduce some error.
	
\begin{figure}[t]
		\centering
		\begin{subfigure}{0.4\textwidth}
			\includegraphics[width=1.0\linewidth]{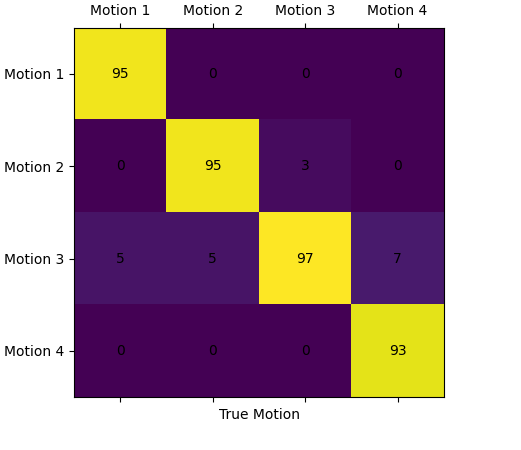}
			\caption{}
		\end{subfigure}
		\begin{subfigure}{0.4\textwidth}
			\includegraphics[width=1.0\linewidth]{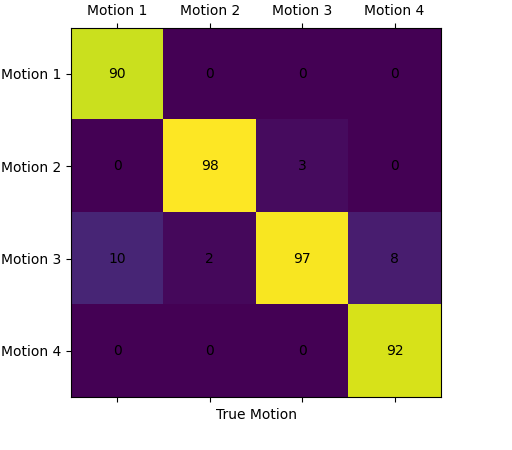}
			\caption{}
		\end{subfigure}
		
		\caption{The confusion matrices for (a) Case 2 and (b) Case 3. The gestures are zero-indexed and thus gesture four represents any gesture that does not match the other three.}
		\label{cm1}
	\end{figure}

	We observed that the model began to over-train at a relativity low number of epochs, which is commonly associated with an over-convolved model or optimization error. This is likely due to the periodicity of the motions which yielded errors between motions 2 and 3 as the general sinusoidal motion of these motions can dominate the more subtle features in activity 2. 
	This may be alleviated when monitoring motion that more accurately replicates daily human motion that has less periodicity; nonetheless, despite this aspect, the classification performance is consistently high.
	Furthermore, more refined micro-Doppler responses such as those described could yield the ability to recognize more complex and multiple motions.
			The confusion matrices for Cases 2 and 3 are shown in Fig. \ref{cm1}.

	\begin{table}[]
		\caption{Classification Accuracy}
		\centering
		\begin{tabular}{lccc}
			\hline\hline \\ [-2mm]
			\textbf{Dataset:}  & \textbf{All Classes}  & \textbf{Lowest Accuracy}\\ 
			\hline \\ [-2mm]
			Case 1: & 99.8\% & 100\% \\
			Case 2: & 95\% & 93\%\\
			Case 3: & 94\% & 90\%\\ [1mm]
			\hline         
		\end{tabular}
		\label{res1}
	\end{table}

	\section{Conclusion}
	
	In this paper, we have shown that a narrow-band harmonic tag combined with a harmonic micro-Doppler radar can produce distinct and identifiable harmonic micro-Doppler spectrogram responses that can be used for motion classification.
	Using a CNN classifier, high accuracy classification can be obtained even for groups of untrained participants. 
	This work demonstrates that detection and classification of held objects in living spaces is viable with radar techniques. 
	The frequencies used in this work furthermore suggest that similar concepts could be implemented in future WiFi systems that use similar frequencies, thereby supporting new applications in human-computer interfacing, IoT, and home health without significant hardware developments. The simplicity of the narrow-band harmonic tags furthermore suggests the feasibility of a low cost solution in future wireless sensing systems.

	\bibliographystyle{IEEEtran}
	\bibliography{nb}

\end{document}